\pdfoutput=1
\documentclass[11pt,twoside,a4paper,cmspaper,final,collab]{cms-tdr}
\def\svnVersion{cd0e5f8}\def\svnDate{2022/06/22}\def\cmsCernNoTag{CERN-EP-2021-259}\def\cmsCernDate{\today}\def\cmsMessage{Published in Physical Review Letters as \href{http://dx.doi.org/10.1103/PhysRevLett.128.252301}{\doi{10.1103/PhysRevLett.128.252301}.}}
\begin{document}\cmsNoteHeader{HIN-20-004}

\providecommand{\PGU}{\ensuremath{\Upsilon}\xspace}
\newcommand{\PGUa}{\ensuremath{\PGU\text{(1S)}}\xspace}
\newcommand{\PGUn}{\ensuremath{\PGU\text{(nS)}}\xspace}

\newcommand{\ptvis}{\ensuremath{p_\text{T}^{\mu\mu\mu}}\xspace}
\newcommand{\yvis}{\ensuremath{y^{\mu\mu\mu}}\xspace}
\newcommand{\mvis}{\ensuremath{m_{\mu\mu\mu}}\xspace}

\providecommand{\pp}{\ensuremath{\Pp\Pp}\xspace}
\providecommand{\PbPb}{\ensuremath{\text{PbPb}}\xspace}

\newcommand{\sqrts}{\ensuremath{\sqrt{s}}\xspace}
\newcommand{\sqrtsnn}{\ensuremath{\sqrt{s_{_{\mathrm{NN}}}}}\xspace}

\providecommand{\CASCADE} {{\textsc{cascade}}\xspace}
\providecommand{\HYDJET} {{\textsc{hydjet}}\xspace}
\providecommand{\BCVEGPYv} {{\textsc{bcvegpy}2.2}\xspace}
\providecommand{\BCVEGPY} {{\textsc{bcvegpy}}\xspace}
\providecommand{\PYTHIAv} {{\textsc{pythia}8}\xspace}
\providecommand{\combine} {{\textsc{combine}}\xspace}

\newcommand{\RAABc}{\ensuremath{R_{\mathrm{AA}}(\PBpc)}\xspace}
\newcommand{\RAA}{\ensuremath{R_{\mathrm{AA}}}\xspace}

\ifthenelse{\boolean{cms@external}}{\providecommand{\suppMaterial}{the supplemental material~\cite{suppMat}}}{\providecommand{\suppMaterial}{Appendix~\ref{app:suppMat}}}

\cmsNoteHeader{HIN-20-004}
\title{Observation of the \texorpdfstring{\PBpc}{Bc+} meson in \texorpdfstring{\PbPb and \pp}{PbPb and pp} collisions at \texorpdfstring{$\sqrtsnn=5.02\TeV$}{sqrt(s[NN]) = 5.02 TeV} and measurement of its nuclear modification factor}

\author*[cern]{The CMS Collaboration}
\date{\today}

\abstract{
The \PBpc meson is observed for the first time in heavy ion collisions. Data from the CMS detector are used to study the production of the \PBpc meson in lead-lead (\PbPb) and proton-proton (\pp) collisions at a center-of-mass energy per nucleon pair of $\sqrtsnn = 5.02\TeV$, via the $\PBpc \to (\PJGy \to \PGmp \PGmm) \PGmp \PGnGm$ decay. The \PBpc nuclear modification factor, derived from the \PbPb-to-\pp ratio of production cross sections, is measured in two bins of the trimuon transverse momentum and of the \PbPb collision centrality. 
The \PBpc meson is shown to be less suppressed than quarkonia and most of the open heavy-flavor mesons, suggesting that effects of the hot and dense nuclear matter created in heavy ion collisions contribute to its production.
This measurement sets forth a promising new probe of the interplay of suppression and enhancement mechanisms in the production of heavy-flavor mesons in the quark-gluon plasma.
}

\hypersetup{%
pdfauthor={CMS Collaboration},%
pdftitle={Observation of the Bc+ meson in PbPb and pp collisions at sqrt(s[NN])=5.02 TeV and measurement of its nuclear modification factor},%
pdfsubject={CMS},%
pdfkeywords={CMS, heavy ion, quarkonia, b mesons, Bc meson}}

\maketitle 

The very high temperatures and densities reached in high-energy heavy ion collisions allow quarks and gluons to form a deconfined state of matter, often referred to as the quark-gluon plasma (QGP). Among the various signatures of the QGP formation, the suppression of heavy quarkonia (e.g.  \PJGy or \PGU mesons) due to the screening of the heavy-quark potential has been intensely studied since its proposal by Matsui and Satz~\cite{Matsui:1986dk}. The strong \PJGy suppression observed in heavy ion collisions at the Super Proton Synchrotron~\cite{Abreu:2000ni} and Relativistic Heavy Ion Collider~\cite{Adare:2006ns} was indeed qualitatively consistent with color screening effects. However, lead-lead (PbPb) collisions at the Large Hadron Collider (LHC) reach higher temperatures, but show less \PJGy suppression~\cite{Abelev:2013ila}. This observation is interpreted as arising from the formation of bound states of charm quarks originating from different hard scatterings, a mechanism referred to as recombination~\cite{BraunMunzinger:2000px,Thews:2000rj}. 
By  contrast, the bottomonium LHC data show no evidence for recombination, consistent with the relatively small \PQb-quark production cross section. In addition, the suppression of \PGUn states is sequential (more suppression for smaller binding energies)~\cite{Chatrchyan:2012lxa}, as expected from color screening.

While understanding the effects of the QGP on heavy-quark bound states is progressing~\cite{Andronic:2015wma}, quantifying them still remains a key challenge. The \PBpc meson is composed of a charm quark and a bottom antiquark, and is intermediate in size and binding energy between the \PJGy and \PGUa mesons~\cite{Liu:2012tn}. Since it bridges the gap between charmonia and bottomonia, the modification of \PBpc yields contributes to probing heavy quark interactions at extreme temperatures.

The recombination of heavy quarks can also affect the \PBpc yield. The production rates of \PJGy and \PBpc mesons from this process should scale quadratically and linearly, respectively, with the large number of charm quarks in the QGP, up to hundreds in central \PbPb collisions. 
However, as the \PBpc production cross section in proton-proton (\pp) scatterings is small, owing to the dominant production mechanism requiring both $\bbbar$ and $\ccbar$ pairs, the enhancement of \PBpc production in heavy ion collisions may greatly exceed that of the \PJGy meson~\cite{Schroedter:2000ek,Liu:2012tn}. These recombination effects would manifest more strongly at low transverse momentum (\pt)~\cite{Liu:2012tn}. 

For $\pt \gg m(\PBpc)$, \PBpc mesons are produced predominantly via heavy-quark fragmentation~\cite{Berezhnoy:1997fp,Chang:2015qea}, and are therefore sensitive to the energy loss of a massive color triplet charge in the QGP -- possibly causing the suppression observed for other \PB mesons~\cite{Sirunyan:2017oug,Sirunyan:2018zys}, \PJGy mesons from \PB decays~\cite{Sirunyan:2017isk}, and \PD mesons~\cite{Sirunyan:2017xss,Acharya:2018hre}. Conversely, the modification of prompt \PJGy meson production for $\pt \gg m(\PJGy)$~\cite{Sirunyan:2017isk,Aaboud:2018quy} probes the energy loss of a massive color octet state~\cite{Arleo:2017ntr}. Therefore, comparing the \PBpc yield with that of other heavy flavor mesons at large \pt can probe both the mass dependence of energy loss (from a possible dead-cone effect~\cite{Dokshitzer:2001zm}) and its color charge dependence.

The \PBpc meson was first observed in proton-antiproton collisions at the Tevatron in the $\PBpc \to \PJGy \ell^+ \PGn_{\ell}$ decay mode~\cite{Abe:1998wi}. Its ground and excited states were then studied in \pp collisions at the LHC~\cite{Aad:2014laa,Khachatryan:2014nfa,Aaij:2014jxa,Sirunyan:2019osb,Aaij:2019ldo}. In this Letter, the first observation of \PBpc mesons produced in heavy ion collisions is reported, and their cross sections are measured and compared in \PbPb and \pp collisions. 
The data were collected with the CMS detector in 2017 for \pp and in 2018 for \PbPb collisions at the same center-of-mass energy per nucleon pair, $\sqrtsnn = 5.02\TeV$, corresponding to integrated luminosities of 302\pbinv and 1.61\nbinv, respectively. The signal is reconstructed from the three muons in the $\PBpc \to (\PJGy \to \PGmp \PGmm) \PGmp \PGnGm$ decay mode. While this mode features a neutrino that prevents a full reconstruction of the decay, it has a much larger branching fraction than neutrinoless decay channels~\cite{Aaij:2014jxa}. In this Letter, charge-conjugate states are implied, and the quoted cross sections correspond to the sum of \PBpc and \PBmc mesons.

The results are presented in two kinematic regions that are defined in terms of the vector sum of the three muon momenta, and whose limits are chosen based on the single-muon acceptance of the CMS apparatus: a low-\ptvis bin, $6 < \ptvis < 11\GeV$ with rapidity $1.3 < \abs{\yvis} < 2.3$, and a high-\ptvis bin, $11 < \ptvis < 35\GeV$ with $\abs{\yvis} < 2.3$. In simulations, the trimuon \pt is, on average, about $15\%$ smaller than the \PBpc \pt. 
In \PbPb collisions, the analysis is performed in the 0--90\% centrality range, where centrality refers to the fraction of the inelastic nucleus-nucleus cross section, with lower values denoting a larger overlap of the nuclei~\cite{LoizidesImprovedMC}. The results integrated over the two kinematic regions are also presented, separated in the centrality ranges 0--20 and 20--90\%. To reduce potential biases, the analysis was performed in a ``blind'' way: the algorithms and selection procedures were finalized and formally approved using a quarter of the \PbPb data, before examining the entire sample. Tabulated results are provided in a HEPData record~\cite{hepdata}.

The central feature of the CMS apparatus~\cite{Chatrchyan:2008zzk} is a superconducting solenoid providing a magnetic field of 3.8\unit{T}. Within the solenoid volume are a silicon pixel and strip tracker, a lead tungstate crystal electromagnetic calorimeter, and a brass and scintillator hadron calorimeter, each composed of a barrel and two endcap sections. Hadron forward calorimeters extend the pseudorapidity coverage to $3 < \abs{\eta} < 5$, and the sum of the transverse energy deposited in them is used to estimate the collision centrality. Muons are detected in gas-ionization detectors embedded in the steel flux-return yoke outside the solenoid and covering $\abs{\eta} < 2.4$. Muons with $\pt>1.2$ and $3.3\GeV$ are reconstructed in the endcap and barrel regions, respectively~\cite{Sirunyan:2018fpa}. For $\pt=1.2\GeV$ muons in the endcaps, the transverse and longitudinal impact parameter resolutions are 150 and 400\micron, respectively, which improve to 20 and 40\micron for $\pt=10\GeV$ muons in the barrel~\cite{Chatrchyan:2014fea}. 

Events of interest are selected using a two-tiered trigger system~\cite{Khachatryan:2016bia}. The first level, composed of custom hardware processors, uses information from the calorimeters and muon detectors~\cite{Sirunyan:2020zal}. The high-level trigger consists of a farm of processors running a fast version of the full event reconstruction software. The events used in this analysis were selected by triggers designed to collect all events containing a \PJGy meson, hence requiring two muons, without \pt requirements. Loose criteria on the single-muon quality and the dimuon mass and opening angle are also applied in the \PbPb high-level trigger, as in Refs.~\cite{CMS:2020efs,CMS:2021puf}. 

Monte Carlo (MC) simulations are used for various signal and background studies, and for estimating the acceptance and efficiency of the reconstruction, triggering, and selection. The \PBpc mesons are generated with \BCVEGPYv~\cite{Chang:2015qea}, while their decays are handled with \textsc{evtgen}1.3~\cite{Lange:2001uf}. The underlying event is generated with \PYTHIA{8.212}~\cite{Sj_strand_2015}, tune CP5~\cite{Sirunyan_2020}.  \PYTHIA{8} is also used to generate prompt and nonprompt (from \PB meson decays) \PJGy background samples. To simulate \PbPb collisions, the generated events are embedded into simulated \PbPb collisions created using \textsc{hydjet1.8}~\cite{Lokhtin_2006}. All samples are passed to \GEANTfour~\cite{Agostinelli:2002hh} to simulate the detector response, and then reconstructed with the same software as the collision data.

The $\PBpc \to (\PJGy \to \PGmp \PGmm) \PGmp \PGnGm$ decay features three muons originating from the same displaced vertex, an opposite-sign muon pair consistent with the \PJGy mass, and a trimuon invariant mass \mvis between $m_{\PJGy}+m_{\PGm} \simeq3.2\GeV$ and $m_{\PBpc} \simeq6.3\GeV$. Three main background sources can mimic this topology. \emph{Fake \PJGy} events arise when neither of the opposite-sign muon pairs originate from a \PJGy decay. It is estimated by summing the trimuon mass distributions obtained in the lower and higher dimuon mass sidebands of \PJGy candidates. The second category (\emph{\PB decays}) comes from \PQb-hadrons (excluding \PBpc) decaying to a true \PJGy meson associated with a muon (usually a misidentified hadron) from the same \PQb-hadron decay. It is estimated via simulation, where the \pt spectrum is corrected using nonprompt \PJGy production measurements~\cite{Sirunyan:2017isk}. Its normalization is unconstrained to cover a possible mismodeling of the muon misidentification rate. 
The third contribution (\emph{\PJGy + random $X$}) combines a true \PJGy meson with a muon candidate (usually an uncorrelated misidentified hadron) from another decay. It is estimated in data by rotating the momentum and decay vertex of \PJGy candidates around the collision vertex before associating them with third muon candidates. Several azimuthal rotation angles (excluding the vicinity of the original \PJGy meson) are used, with or without inverting rapidity. In \PbPb collisions, the associated muons are mostly uncorrelated with the \PJGy meson, so the distributions from various rotation angles are identical (within statistical uncertainties) and averaged, with a data-derived normalization (fixed in the fit). In \pp collisions, significant residual $\PJGy-\PGm$ correlations lead to different distributions for different rotation angles, which is accounted for by considering various mixes of these distributions. 

The offline selection includes the same event-level and single-muon identification criteria as in Refs.~\cite{CMS:2020efs,CMS:2021puf}. Loose kinematic acceptance criteria are applied to the muon candidates, matching the efficient region for the two triggering muons, and even looser for the third one. At least one of the two opposite-sign dimuon combinations must have an invariant mass in the \PJGy peak region, or in the sidebands used for background estimation. 
The sideband and peak regions are both asymmetric to account for radiative tails, and are separated by small gaps. The total sideband width equals that of the peak region, from 180 to 260\MeV depending on the muon pseudorapidity (which affects the mass resolution). 
For the trimuons having two opposite-sign dimuons in the studied mass regions (5-6\% of the overall sample), the two corresponding trimuon candidates are kept, weighted by the probability of the chosen dimuon to be a true \PJGy meson. This probability is extracted from the dimuon mass distribution from events with only one \PJGy candidate in the signal or sideband regions.

Requirements are also set on the probability of the trimuon vertex fit, the significance of its displacement from the collision vertex, the angle between the trimuon momentum and the segment joining the collision and trimuon vertices, the invariant mass corrected for the momentum of the neutrino transverse to the \PBpc momentum direction, and the sum of the angular separations $\Delta R = \sqrt{\smash[b]{(\Delta \eta)^2 + (\Delta \phi)^2 }}$ between the three muon pairs. 

After the selection, the simulated signal and the three background samples are used to train a boosted decision tree (BDT) using the TMVA package~\cite{Hocker:2007ht}. This combines and optimizes the discriminating power of eight variables: the five discussed in the previous paragraph, the imbalance between the \pt of the \PJGy and of the third muon, the ratio of the $\Delta R$ of the \PJGy muons to the sum of the $\Delta R$ values from the other two dimuon combinations, and the significance of the displacement from the collision vertex for the non-\PJGy muon.

Candidates with very low values of the resulting discriminant BDT variable (hence very high background probability) are rejected, losing only 0.1\% in signal efficiency. For each analysis bin, low, medium, and high BDT intervals are set to contain about 25, 40, and 35\%, respectively, of the expected signal. The first and last intervals are dominated by background and signal, respectively. A binned likelihood fit of the \pp or \PbPb trimuon mass distributions provides the signal yields. Using \textsc{roofit}~\cite{Verkerke:2003ir}, templates from the signal and the three backgrounds are simultaneously fitted in the three BDT intervals, and in either two kinematic bins, two centrality bins, or the whole kinematic range.
The BDT distribution of the sum of the fitted templates is checked against that of data, and, in \pp collisions, corrected before re-running the template fit.

\begin{figure*}[ht!]
\centering
\includegraphics[width=\textwidth]{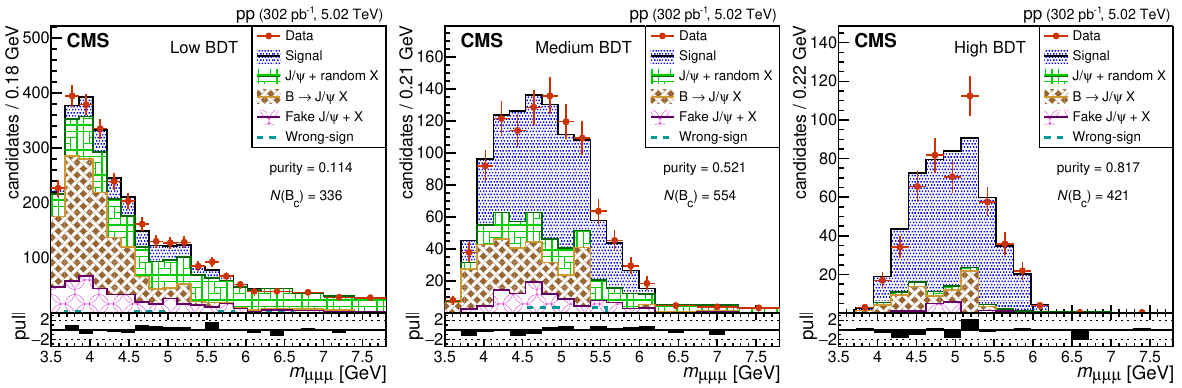}
\includegraphics[width=\textwidth]{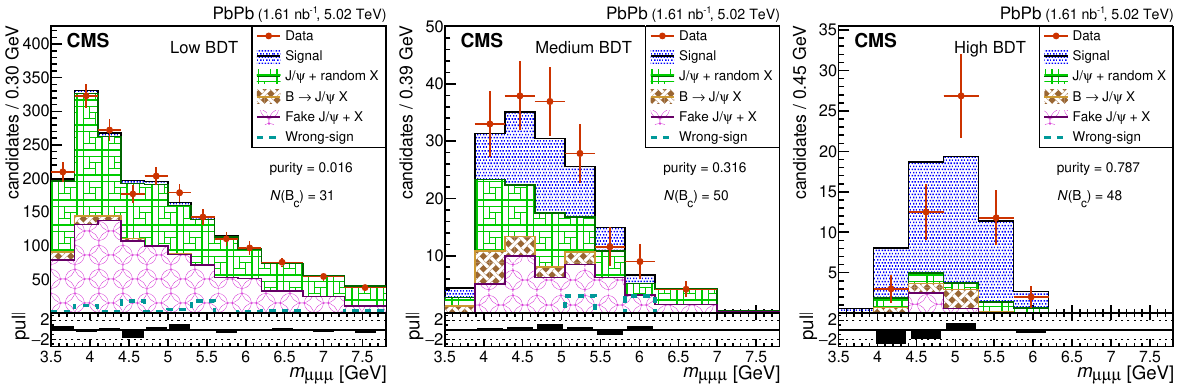}
\caption{Template fit of the trimuon mass distributions in the three BDT bins, for the \pp (top row) and \PbPb (bottom row) data samples integrated over the two studied kinematic regions. The lower panels show the pull between the data and the fitted distributions.
 }
\label{fig:PbPbFit}
\end{figure*}

The results of the fits in the three BDT intervals and integrated over the two kinematic regions are shown for \pp and \PbPb collisions in Fig.~\ref{fig:PbPbFit}. In each BDT bin, 
the signal purity and the measured yield, $N(\PBc)$, are given. The wrong-sign distributions, containing three same-sign muons in data, are superimposed to illustrate that the purely combinatorial background is easily rejected. The normalizations of the fake \PJGy sample, and of the \PJGy + random $X$ sample in \PbPb collisions, are provided by the data. In \PbPb collisions, the \PJGy + random $X$ and fake \PJGy backgrounds are dominant. In \pp collisions, the region above the \PBpc mass strongly constrains the \PJGy + random $X$ contribution, and the remaining background comes from \PB decays and fake \PJGy events. 

The signal yields extracted from the fit are corrected for the acceptance and efficiency of the reconstruction, triggering, and selection. These are calculated in each analysis bin using the simulated signal trimuons. The simulated efficiencies of single muon reconstruction, identification and triggering are corrected by a tag-and-probe method using the \PJGy resonance, similarly to Refs.~\cite{Sirunyan:2017isk,CMS:2020efs}. The acceptance and efficiency are evaluated iteratively by first performing the \mbox{\ptvis-differential} analysis using the original simulation. The resulting corrected yields are fitted to correct the \ptvis spectrum of the simulation before a second run of the analysis. This \ptvis spectrum is then corrected again based on the second-step results, notably improving upon the initial acceptance and efficiency estimation. 

The corrected yields are divided by the \pp integrated luminosity~\cite{CMS-PAS-LUM-19-001} or by its \PbPb equivalent, the number of minimum bias \PbPb hadronic collisions $N_{\mathrm{MB}}$ times the nuclear overlap function $T_{\PbPb}$ from Ref.~\cite{LoizidesImprovedMC}. 
The \PbPb-to-\pp ratio of these \pp-equivalent normalized yields then provides the nuclear modification factor, \RAA. In case of no modification by the medium, \RAA is expected to be equal to unity. 

Uncertainties arise from different sources: \emph{statistical}, \emph{background} (shapes and normalizations), choice of the \emph{fit method}, \emph{muon efficiency}, \PBpc \emph{kinematics} (acceptance and efficiency), \emph{contamination} from other \PBpc decays, and overall \emph{normalization}. The fit uncertainties, ranging from 5 to 9\% in \pp and 17 to 31\% in \PbPb collisions, include the purely \emph{statistical} and the \emph{background} uncertainties. The latter are implemented via nuisance parameters allowing variations of the trimuon mass templates, such as controlling their statistical uncertainties with the Barlow-Beeston procedure~\cite{Conway:2011in}, varying the fake \PJGy background between the lower and higher dimuon sideband, or varying the rotation angles in the \PJGy + random $X$ background. Variations of the \emph{fit method} are also considered, such as changing the \mvis or BDT bin size, neglecting the low-BDT bin, using a BDT variable whose \mvis dependence is subtracted, or regularizing the low-statistics templates instead of using the Barlow-Beeston procedure. The resulting uncertainty remains below 7\% (12\%) in \pp (\PbPb) collisions. The uncertainty from the tag-and-probe derived \emph{muon efficiency} corrections is 2--5\%. 

Since the \PBpc \emph{kinematic} distributions are not precisely known, acceptance and efficiency corrections are recalculated 1500 times with \ptvis spectra fitted on variations of the measured \ptvis-differential yields within the above-mentioned uncertainties. For the \ptvis-integrated results, the root-mean-square (RMS) of the varied acceptance and efficiency corrections, of order 7 and 24\% for \pp and \PbPb collisions, respectively, is used as the systematic uncertainty related to the \PBpc kinematics. For the \ptvis dependence, these variations are correlated with the other uncertainty sources, so the combined uncertainty is assessed as the RMS of the varied corrected yields. 
The correlation between the variations of the spectrum and of the acceptance and efficiency is small or negative for the \PbPb high-\ptvis bin and for both \pp \pt bins, so that the uncertainties with or without this systematic effect are similar. 
This correlation is large and positive for the \PbPb low-\ptvis bin, inducing an additional 12--31\% uncertainty. The uncertainty in the \PbPb-to-\pp ratio is the RMS of the ratios of the relevant varied quantities.
	
The \emph{contamination} from other \PBpc decays, such as $\PBpc \to \PJGy (\PGtp \to \PGmp X)\, \PGn_{\PGt}$ or $\PBpc \to (\ccbar \to \PJGy X)\, \PGmp \PGn_{\PGm}$, where $X$ denotes any decay product(s), is estimated to be below 4.5\%, and to have largely cancelling \pp and \PbPb contributions. The overall \emph{normalization} uncertainty arising from the luminosity and centrality determination is 1.9--3.8\%. The leading uncertainties in the \ptvis-differential and \ptvis-integrated measurements are from the fit and the \PBpc kinematics, respectively. 

\begin{figure*}[htb!]
\centering
 	\includegraphics[width=0.32\textwidth]{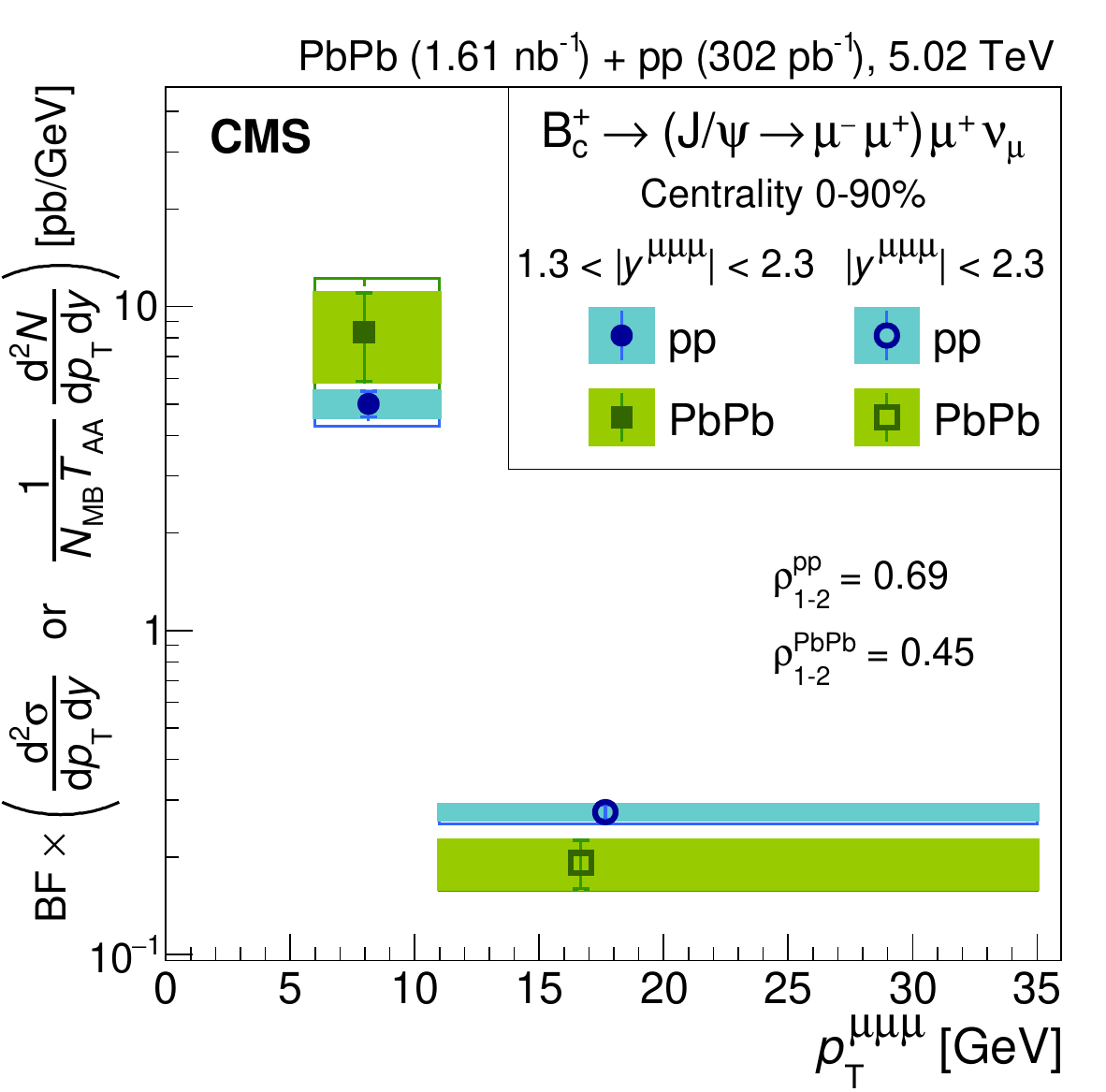}
	\includegraphics[width=0.32\textwidth]{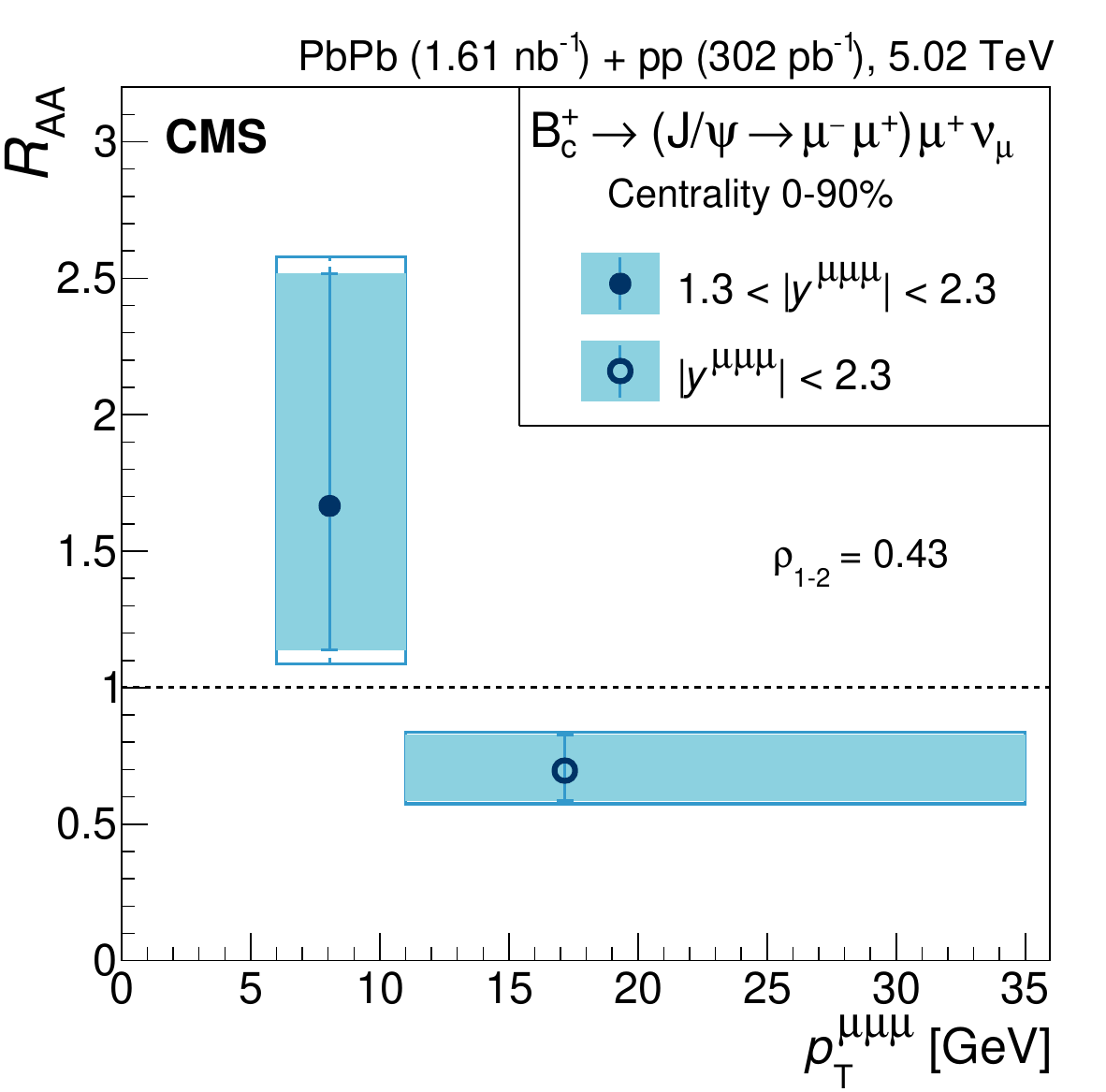}
	\includegraphics[width=0.32\textwidth]{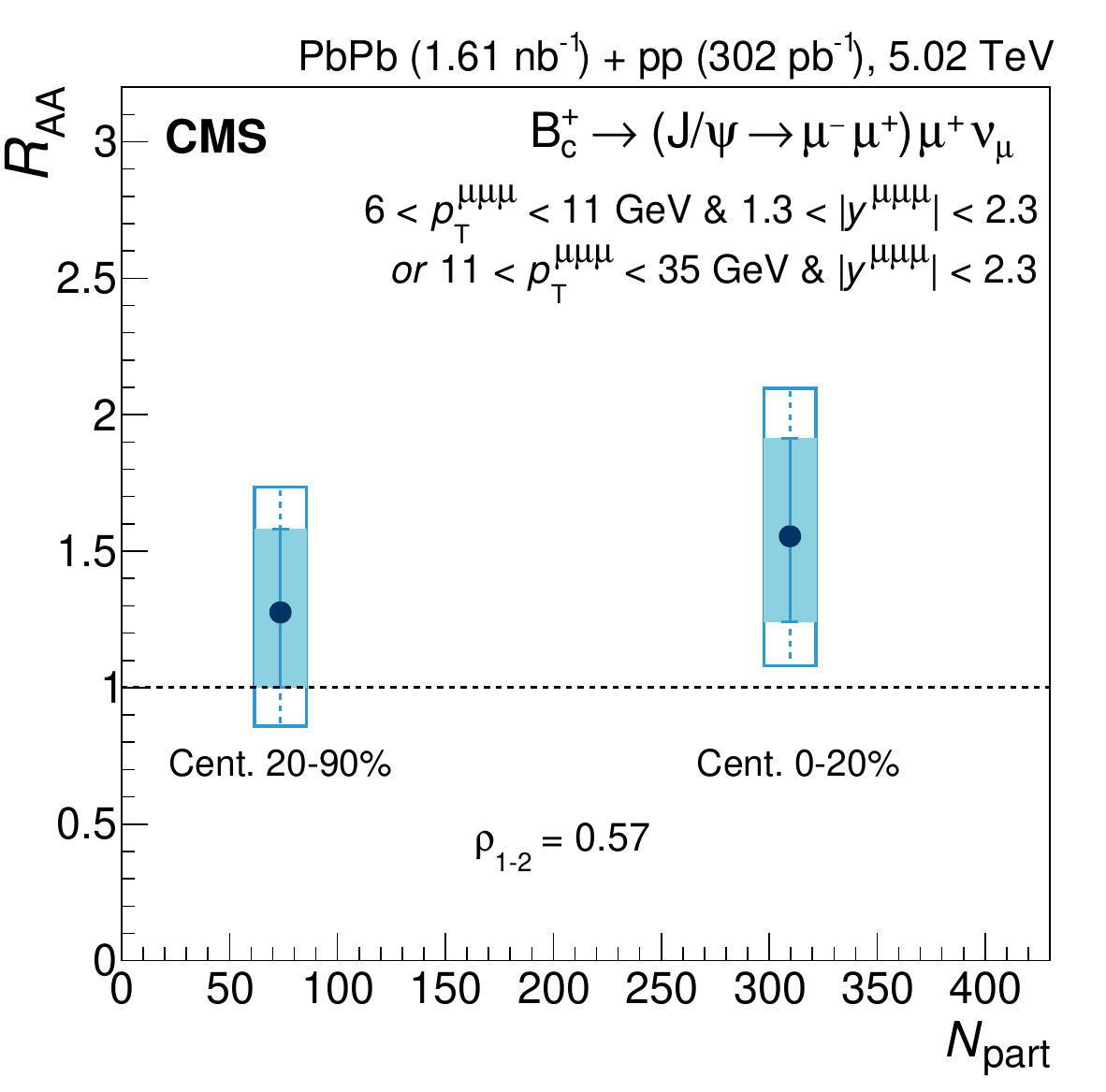}
\caption{
\PBpc meson production in \pp and \PbPb collisions. Left: The \pp-equivalent cross-section times branching fraction of the studied decay in \pp and \PbPb collisions. Middle and right: The nuclear modification factor in \ptvis bins (middle), and in centrality bins integrated over the studied kinematic range (right). See the text for details. 
}
\label{fig:RAA}
\end{figure*}

The significance of the \PBpc signal in \PbPb collisions, calculated from the fit likelihood ratio and including the fit method uncertainty, is well above 5 standard deviations. 
The left panel of Fig.~\ref{fig:RAA} shows the measured \PBpc meson \ptvis-differential cross sections in \pp and (\pp-equivalent) \PbPb collisions. 
The two bins of the trimuon \pt correspond to different rapidity ranges. The markers of the \ptvis bins are placed according to the Lafferty--Wyatt prescription~\cite{Lafferty:1994cj}.
The bin-to-bin correlation factor $\rho_\text{1-2}$ is also displayed. The filled and empty rectangles show the fit and  total uncertainties, respectively. 
The ratio between the low-\ptvis and high-\ptvis regions is $18.2^{+1.3}_{-2.1}$ in \pp data and 24.1 in the \BCVEGPYv simulation, suggesting that the latter overestimates the spectrum steepness.

The other panels of Fig.~\ref{fig:RAA} show the \PBpc nuclear modification factor, \ie, the ratio of the (\pp-equivalent) \PbPb to \pp cross sections, as a function of \ptvis (middle) and of centrality (right). 
The markers of the \ptvis bins are placed at the average of their values for \pp and \PbPb collisions, while the centrality bin markers are placed at the minimum bias average number of participants $N_{\text{part}}$.
The filled and empty rectangles, respectively, show the bin-to-bin-uncorrelated and total uncertainties, such that the uncertainty in the difference of the two bins is the quadratic sum of uncorrelated uncertainties. 

In the high-\ptvis region, the \PBpc shows a moderate suppression, while the low-\ptvis modification factor stands above unity and above the high-\ptvis region, respectively, by 1.2 and 1.8 standard deviations, consistent with an enhancement of the integrated production and a softening of the \pt spectrum in the QGP\@. No significant variation is observed as a function of centrality. As shown in \suppMaterial, except for the \PBzs meson~\cite{Sirunyan:2018zys}, other heavy mesons in these \pt ranges typically show more suppression than our measurement~\cite{Abelev:2013ila,Chatrchyan:2012lxa,Sirunyan:2017oug,Sirunyan:2017isk,Sirunyan:2017xss,Acharya:2018hre}, which may indicate that heavy-quark recombination is a significant \PBpc production mechanism in the QGP\@. A study based on Ref.~\cite{Yao:2020xzw}, ignoring the recombination of \PBpc excited states and possibly underestimating initial correlations, predicts an \RAABc about an order of magnitude smaller than our measurement.

In summary, the first observation of the \PBpc meson in heavy ion collisions is presented, using the $\PBpc \to (\PJGy \to \PGmp \PGmm) \PGmp \PGn_{\PGm}$ decay. The production cross sections in lead-lead and proton-proton collisions and the  nuclear modification factor derived from their ratio are measured in two bins of the trimuon transverse momentum, and in two ranges of the heavy-ion centrality. This unique bottom-charm state can help disentangle the enhancement (possibly dominant in central events at low \pt) and suppression (dominant at high \pt) mechanisms at play in the evolution of heavy quarks through the quark-gluon plasma. 

\begin{acknowledgments}
	We congratulate our colleagues in the CERN accelerator departments for the excellent performance of the LHC and thank the technical and administrative staffs at CERN and at other CMS institutes for their contributions to the success of the CMS effort. In addition, we gratefully acknowledge the computing centers and personnel of the Worldwide LHC Computing Grid and other centers for delivering so effectively the computing infrastructure essential to our analyses. Finally, we acknowledge the enduring support for the construction and operation of the LHC, the CMS detector, and the supporting computing infrastructure provided by the following funding agencies: BMBWF and FWF (Austria); FNRS and FWO (Belgium); CNPq, CAPES, FAPERJ, FAPERGS, and FAPESP (Brazil); MES and BNSF (Bulgaria); CERN; CAS, MoST, and NSFC (China); MINCIENCIAS (Colombia); MSES and CSF (Croatia); RIF (Cyprus); SENESCYT (Ecuador); MoER, ERC PUT and ERDF (Estonia); Academy of Finland, MEC, and HIP (Finland); CEA and CNRS/IN2P3 (France); BMBF, DFG, and HGF (Germany); GSRI (Greece); NKFIA (Hungary); DAE and DST (India); IPM (Iran); SFI (Ireland); INFN (Italy); MSIP and NRF (Republic of Korea); MES (Latvia); LAS (Lithuania); MOE and UM (Malaysia); BUAP, CINVESTAV, CONACYT, LNS, SEP, and UASLP-FAI (Mexico); MOS (Montenegro); MBIE (New Zealand); PAEC (Pakistan); MSHE and NSC (Poland); FCT (Portugal); JINR (Dubna); MON, RosAtom, RAS, RFBR, and NRC KI (Russia); MESTD (Serbia); MCIN/AEI and PCTI (Spain); MOSTR (Sri Lanka); Swiss Funding Agencies (Switzerland); MST (Taipei); ThEPCenter, IPST, STAR, and NSTDA (Thailand); TUBITAK and TAEK (Turkey); NASU (Ukraine); STFC (United Kingdom); DOE and NSF (USA).	
\end{acknowledgments}

\bibliography{auto_generated}

\ifthenelse{\boolean{cms@external}}{}{
\clearpage
\appendix
\numberwithin{table}{section}
\numberwithin{figure}{section}
\section{Comparison of the nuclear modification of \texorpdfstring{\PBpc}{Bc+} vs.\ open and hidden heavy flavor mesons\label{app:suppMat}}
\begin{figure}[htb!]
    \centering
         \includegraphics[width=0.49\textwidth]{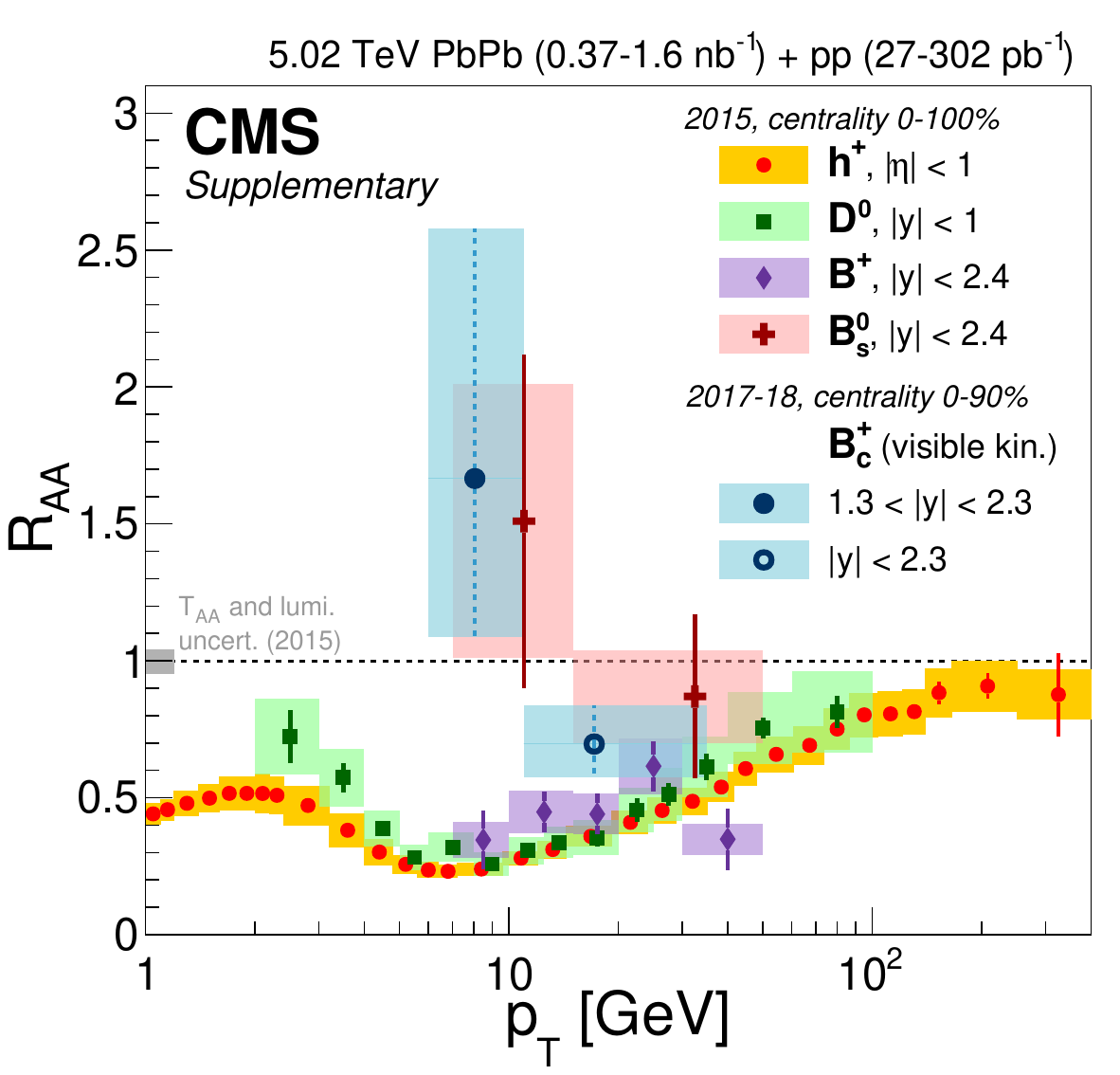}
    \caption{Nuclear modification factor of the \PBpc meson compared to that of light charged hadrons, and \PBp, \PBs and \PDz mesons, as a function of the measured transverse momentum. The total uncertainty is shown for the \PBpc meson, whereas the statistical (bars) and systematic (filled rectangles) uncertainties are shown for the other hadrons.}
    \end{figure}
    
    \begin{figure}[htb!]
    \centering
         \includegraphics[width=0.49\textwidth]{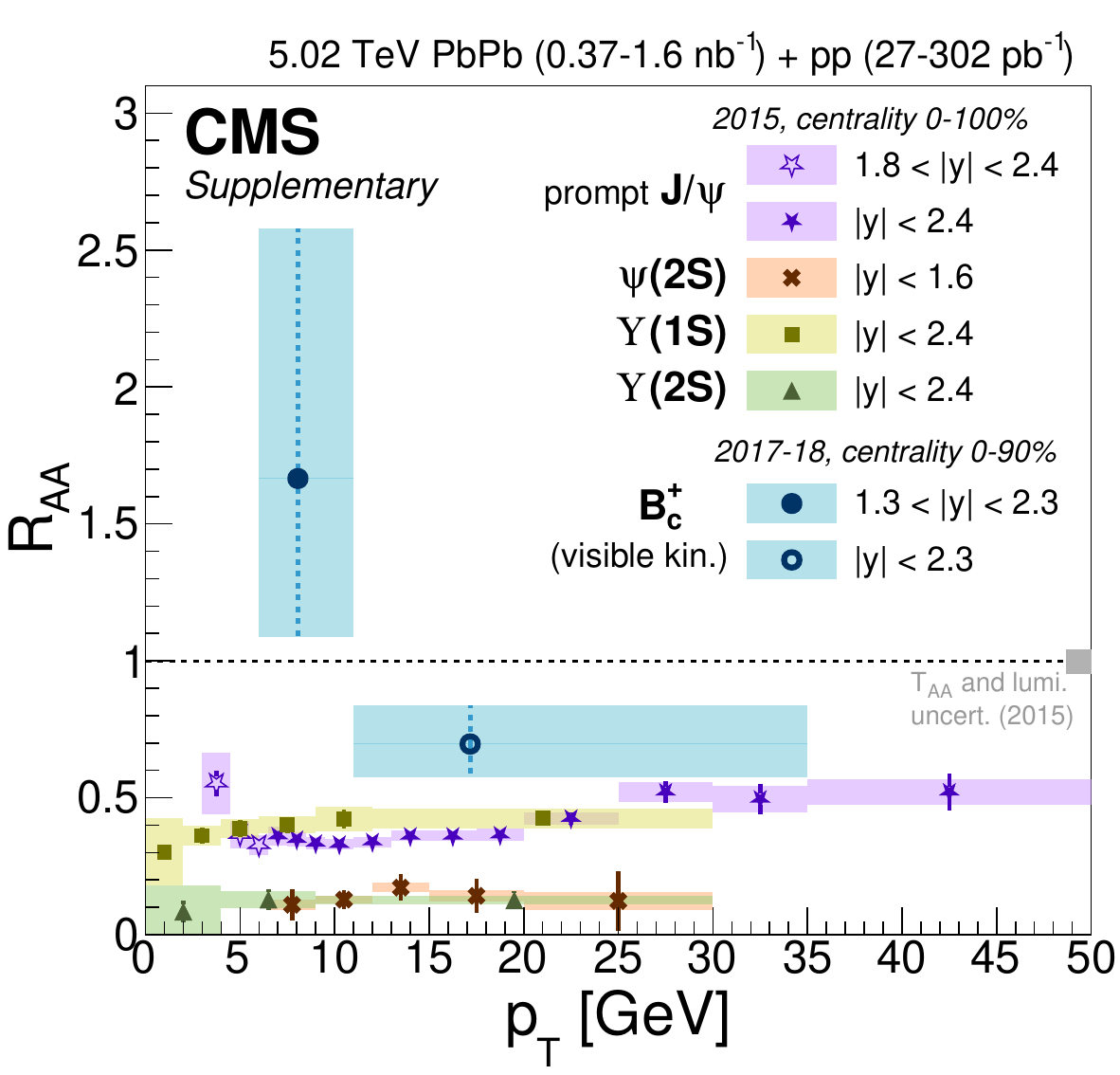}
    \caption{Nuclear modification factor of the \PBpc meson compared to that of the ground and first excited states of charmonia and bottomonia, as a function of the measured transverse momentum. The total uncertainty is shown for the \PBpc meson, whereas the statistical (bars) and systematic (filled rectangles) uncertainties are shown for the other hadrons.}
    \end{figure}
}

\cleardoublepage \section{The CMS Collaboration \label{app:collab}}\begin{sloppypar}\hyphenpenalty=5000\widowpenalty=500\clubpenalty=5000\cmsinstitute{Yerevan~Physics~Institute, Yerevan, 8732Armenia}
A.~Tumasyan
\cmsinstitute{Institut~f\"{u}r~Hochenergiephysik, Vienna, Austria}
W.~Adam\cmsorcid{0000-0001-9099-4341}, J.W.~Andrejkovic, T.~Bergauer\cmsorcid{0000-0002-5786-0293}, S.~Chatterjee\cmsorcid{0000-0003-2660-0349}, M.~Dragicevic\cmsorcid{0000-0003-1967-6783}, A.~Escalante~Del~Valle\cmsorcid{0000-0002-9702-6359}, R.~Fr\"{u}hwirth\cmsAuthorMark{1}, M.~Jeitler\cmsAuthorMark{1}\cmsorcid{0000-0002-5141-9560}, N.~Krammer, L.~Lechner\cmsorcid{0000-0002-3065-1141}, D.~Liko, I.~Mikulec, P.~Paulitsch, F.M.~Pitters, J.~Schieck\cmsAuthorMark{1}\cmsorcid{0000-0002-1058-8093}, R.~Sch\"{o}fbeck\cmsorcid{0000-0002-2332-8784}, D.~Schwarz, S.~Templ\cmsorcid{0000-0003-3137-5692}, W.~Waltenberger\cmsorcid{0000-0002-6215-7228}, C.-E.~Wulz\cmsAuthorMark{1}\cmsorcid{0000-0001-9226-5812}
% ...
\vskip\cmsinstskip
% ...
1:~Also at TU Wien, Wien, Austria\\
% ...
\end{sloppypar}
%%% END EDITABLE REGION %%%
% skeleton_end
\end{document}